\documentclass[aps,pra,graphicx,groupedaddress,reprint,showpacs]{revtex4-1}
\usepackage{amsmath}
\usepackage{amssymb}
\usepackage{graphicx}
\usepackage{hyperref}
\usepackage{tensor}
\usepackage{txfontsb} %most up-to-date times font supported by the arXiv
\usepackage{multirow}
\DeclareMathAlphabet{\mathcal}{OMS}{cmsy}{m}{n} %Keep the nice script-O
\hypersetup{
    colorlinks=true,
    linkcolor=blue,
    citecolor=blue,
    filecolor=blue,
    urlcolor=blue,
}
\usepackage[braket]{qcircuit}
\newcommand{\qslash}{*=<0em>{/} \qw}

\begin{document}

\title{Quantum circuits for qubit fusion}

\author{Jonathan E. Moussa}
\email{godotalgorithm@gmail.com}
\affiliation{Center for Computing Research, Sandia National Laboratories, Albuquerque, New Mexico 87185, USA}

\begin{abstract} 
\bigskip
\centerline{\begin{minipage}{0.79\textwidth}
\ \ \ We consider four-dimensional qudits as qubit pairs and their qudit Pauli operators as qubit Clifford operators.
This introduces a nesting, $C_1^2 \subset C_2^4 \subset C_3^2$,
 where $C_n^m$ is the $n$th level of the $m$-dimensional qudit Clifford hierarchy.
If we can convert between logical qubits and qudits, then qudit Clifford operators are qubit non-Clifford operators.
Conversion is achieved by qubit fusion and qudit fission using stabilizer circuits that consume a resource state.
This resource is a fused qubit stabilizer state with a fault-tolerant state preparation using stabilizer circuits.
\end{minipage}}
\bigskip
\bigskip
\end{abstract}

\maketitle

%S1
\section{Introduction}

%P1.1 - Increasing theoretical interest in qudits, but qubits are an experimental reality
There is increasing theoretical interest in using qudits for quantum information processing \cite{qudit_magic_distillation,qudit_surface_code},
 but most experimental efforts are focused on building qubits.
A natural way to relate these activities is to consider embedding qudits in qubits \cite{other_embedding}.
The simplest invertible embedding is a four-dimensional qudit in two qubits.
Quantum error correction is more complicated for qudits of composite dimension,
 but high-distance surface codes can be constructed for qudits of any dimension \cite{composite_surface_code}.

%P1.2 - Relation between qubit and qudit Clifford operations & fusion/fission operations
This paper develops a useful physical relationship between qubits and qudits that is represented as quantum circuits.
All further use of ``qudit'' implicitly refers to a four-dimensional qudit.
We assume the availability of stabilizer operations with negligible errors on both qubits and qudits.
These operations combine projective measurements in a joint Pauli basis
 with unitary transformations in a joint Clifford group that are able to entangle qubits with qudits.
Superficially, a qudit Clifford operation can be a non-Clifford operation on two underlying qubits.
To achieve actual non-Clifford operations, we need to teleport quantum states between two qubits and a qudit.
We then posit the existence of a circuit element for qubit fusion,
\begin{subequations}
\label{fusion_and_fission}
\begin{equation} \label{fusion_element}
\vcenter{\vbox{\Qcircuit @C=0.5em @R=0.4em {
 \lstick{\ket{x}} & \multigate{1}{F} & \qslash & \rstick{\ket{2y+x}} \qw \\
 \lstick{\ket{y}} & \ghost{F}           &             &
}}} \ \ \ \ \ \ \ \ \ \ \ \ \ \ \ ,
\end{equation}
 and another for the conjugate operation of qudit fission,
\begin{equation}
\vcenter{\vbox{\Qcircuit @C=0.5em @R=0.4em {
 \lstick{\ket{x}} & \qslash & \multigate{1}{F^\dag}  & \rstick{\ket{x \bmod 2}} \qw \\
                       &              & \pureghost{F^\dag}             & \rstick{\ket{ \, \lfloor x/2 \rfloor \, }} \qw
}}}  \ \ \ \ \ \ \ \ \ \ \ \ \ \ \ \ \ \, \, .
\end{equation}
\end{subequations}
The relative orientation of input and output wires  is used to distinguish the inequivalent qubit wires of $F$ and $F^\dag$.

%P1.3 - Code conversion as a distillable resource
To implement $F$ or $F^\dag$ using stabilizer operations, a qudit ancilla state must be consumed.
For logical qubits and qudits encoded in different quantum codes, this state is effectively a resource for code conversion.
It can be distilled and teleported to correct faulty $F$ operations analogous to gate teleportation of non-Clifford operations \cite{gate_teleportation}.
This combines aspects of code conversion \cite{conversion} and resource state distillation \cite{distillation}
 to implement an unconventional but universal set of quantum operations.

%P1.4 - Note about notation
We use standard quantum circuit notation \cite{quantum_circuit} throughout the paper with modifications to accommodate qudits.
Qudits are labeled by a slash on the left end of the wire.
Pauli and Clifford gates on qudit wires denote the corresponding qudit operations.
An operation controlled by a qudit in the state $| x \rangle$ is applied $x$ times.
All state preparation and measurement is restricted to the qubit and qudit computational basis.

%S2
\section{Stabilizer operations}

%P2.1 - Assume knowledge of the qubit case
To enable a concise presentation of results, we assume that the reader is familiar with stabilizer operations on qubits that
 map between elements of the Pauli group using elements of the Clifford group \cite{quantum_circuit}.
The Pauli group is generated by an $i$ phase factor and an $X$ and $Z$ operator for each qubit,
 and the Clifford group is generated by an $\omega = \exp(i \pi / 4)$ phase factor,
 a controlled-\textsc{not} (\textsc{cnot}) operation between qubit pairs, and a Hadamard ($H$) and phase ($S$) operation for each qubit.
 
%P2.2 - Summarize the qudit case
We use the standard extension of Pauli and Clifford group structure to qudits \cite{qudit_stabilizer}
 with notation similar to the qubit case.
The Pauli group is still generated by a phase factor and an $X$ and $Z$ operator for each qudit,
 but the phase is now $\omega$ and the Pauli operator algebra on a qudit is summarized by
\begin{subequations}
\begin{align}
 ZX &= i XZ, \\
 Z^{-1} &= Z^\dag = Z^3 , \\
 X^{-1} &= X^\dag = X^3 .
\end{align}
\end{subequations}
Generalizations of \textsc{cnot}, $H$, and $S$ operations still generate the qudit Clifford group up to a global phase \cite{qudit_stabilizer}.
They can be characterized by their action on the qudit $X$ and $Z$ operators,
\begin{subequations}
\label{Clifford_action}
\begin{align}
\vcenter{\vbox{\Qcircuit @C=0.5em @R=0.4em {
 & \qslash & \gate{X} & \ctrl{1} & \qw \\
 & \qslash & \push{\rule{0em}{1.2em}} \qw         & \targ    & \qw
}}} &=
\vcenter{\vbox{\Qcircuit @C=0.5em @R=0.4em {
 & \qslash  & \ctrl{1} & \gate{X} & \qw \\
 & \qslash & \targ &  \gate{X} & \qw
}}} \\
\vcenter{\vbox{\Qcircuit @C=0.5em @R=0.4em {
 & \qslash & \push{\rule{0em}{1.2em}} \qw & \ctrl{1} & \qw \\
 & \qslash & \gate{X} & \targ    & \qw
}}} &=
\vcenter{\vbox{\Qcircuit @C=0.5em @R=0.4em {
 & \qslash  & \ctrl{1} & \push{\rule{0em}{1.2em}} \qw & \qw \\
 & \qslash & \targ &  \gate{X} & \qw
}}} \\
\vcenter{\vbox{\Qcircuit @C=0.5em @R=0.4em {
 & \qslash & \gate{Z} & \ctrl{1} & \qw \\
 & \qslash & \push{\rule{0em}{1.2em}} \qw         & \targ    & \qw
}}} &=
\vcenter{\vbox{\Qcircuit @C=0.5em @R=0.4em {
 & \qslash  & \ctrl{1} & \gate{Z} & \qw \\
 & \qslash & \targ & \push{\rule{0em}{1.2em}} \qw & \qw
}}} \\
\vcenter{\vbox{\Qcircuit @C=0.5em @R=0.4em {
 & \qslash & \push{\rule{0em}{1.4em}} \qw & \ctrl{1} & \qw \\
 & \qslash & \gate{Z} & \targ    & \qw
}}} &=
\vcenter{\vbox{\Qcircuit @C=0.5em @R=0.4em {
 & \qslash  & \ctrl{1} & \gate{Z^\dag} & \qw \\
 & \qslash & \targ &  \gate{Z} & \qw
}}}
\end{align}
\begin{align}
\vcenter{\vbox{\Qcircuit @C=0.5em @R=0.4em {
 & \qslash & \gate{X} & \gate{H} & \qw
}}} &=
\vcenter{\vbox{\Qcircuit @C=0.5em @R=0.4em {
 & \qslash  & \gate{H} & \gate{Z} & \qw
}}} \\
\vcenter{\vbox{\Qcircuit @C=0.5em @R=0.4em {
 & \qslash & \gate{Z} & \gate{H} & \qw
}}} &=
\vcenter{\vbox{\Qcircuit @C=0.5em @R=0.4em {
 & \qslash  & \gate{H} & \gate{X^\dag} & \qw
}}} \\
\vcenter{\vbox{\Qcircuit @C=0.5em @R=0.4em {
 & \qslash & \gate{X} & \gate{S} & \qw
}}} &=
\vcenter{\vbox{\Qcircuit @C=0.5em @R=0.4em {
 & \qslash  & \gate{S} & \gate{Z} & \gate{X} & \qw
}}} \times \omega \\
\vcenter{\vbox{\Qcircuit @C=0.5em @R=0.4em {
 & \qslash & \gate{Z} & \gate{S} & \qw
}}} &=
\vcenter{\vbox{\Qcircuit @C=0.5em @R=0.4em {
 & \qslash  & \gate{S} & \gate{Z} & \qw
}}}
\end{align}
\end{subequations}
It was initially conjectured that other single-qudit operations would be needed to generate the qudit Clifford group \cite{old_qudit_stabilizer},
 but \textsc{cnot}, $H$, and $S$ were recently proven to be sufficient \cite{qudit_stabilizer}.

%P2.3 - Other notational issues
Except for the phase factor in the $S$ gate ($\omega$ rather than $i$), 
Eq. (\ref{Clifford_action}) is consistent with the qubit case.
$X$, $Z$, $H$, and \textsc{cnot} are Hermitian and order two for qubits
 but are non-Hermitian and order four for qudits.
This necessitates operator algebra rules that distinguish an operator from its conjugate.

%P2.4 - Summarize the rest of the section
We present further details of joint stabilizer operations on qubits and qudits in the next two subsections.
In Sec. \ref{standard_qudit}, we specify the standard representation of qudit Pauli and Clifford operations as qubit operations.
In Sec. \ref{hybrid_Clifford}, we define hybrid Clifford operations between qubits and qudits.

%S2.1 - Clock and shift matrices
\subsection{Standard qudit representation\label{standard_qudit}}

%P2.1.1 - Representation in terms of action on computational basis states
We attribute the standard representation of the qudit Pauli group \cite{qudit_stabilizer}
 to the qudits generated by the $F$ gate in Eq. (\ref{fusion_and_fission}).
It is typically defined by the action of Pauli and Clifford group generators on qudit computational basis states.
$X$ is a ``shift'' operation that increments the basis state by one modulo four,
\begin{subequations}
\label{basis_action}
\begin{equation}
\vcenter{\vbox{\Qcircuit @C=0.5em @R=0.4em {
 \lstick{\ket{x}} & \qslash & \gate{X}  & \rstick{ \ket{ (x+1) \bmod 4}} \qw
}}} \ \ \ \ \ \ \ \ \ \ \ \ \ \ \ \ \ \ \ \ \ \ \ \ \ \ \ .
\end{equation}
$Z$ is a ``clock'' operation that shifts the phase by a power of $i$,
\begin{equation}
\vcenter{\vbox{\Qcircuit @C=0.5em @R=0.4em {
 \lstick{\ket{x}} & \qslash & \gate{Z}  & \rstick{ i^x \ket{x}} \qw
}}}  \ \ \ \ \ \ \ \ \ \ .
\end{equation}
\textsc{cnot} is a modular addition operation on the basis index,
\begin{equation}
\vcenter{\vbox{\Qcircuit @C=0.65em @R=0.8em {
 \lstick{\ket{x}} & \qslash & \ctrl{1} & \rstick{ \ket{x}} \qw \\
 \lstick{\ket{y}} & \qslash & \targ & \rstick{ \ket{(x+y) \bmod 4}} \qw
}}} \ \ \ \ \ \ \ \ \ \ \ \ \ \ \ \ \ \ \ \ \ \ \ \ \ \ \ .
\end{equation}
$H$ is a discrete Fourier transform of the quantum state,
\begin{equation}
\vcenter{\vbox{\Qcircuit @C=0.5em @R=0.4em {
 \lstick{\ket{x}} & \qslash & \gate{H}  & \rstick{ \tfrac{1}{2} \sum_{y=0}^3 i^{xy} \ket{y}} \qw
}}} \ \ \ \ \ \ \ \ \ \ \ \ \ \ \ \ \ \ \ \ \ \ \ .
\end{equation}
$S$ is a phase shift by a power of $\omega$ with a quadratic exponent,
\begin{equation}
\vcenter{\vbox{\Qcircuit @C=0.5em @R=0.4em {
 \lstick{\ket{x}} & \qslash & \gate{S}  & \rstick{ \omega^{x^2} \ket{x}} \qw
}}} \ \ \ \ \ \ \ \ \ \ \ \ \ .
\end{equation}
\end{subequations}
Other suggested qudit Clifford operations \cite{old_qudit_stabilizer} are redundant.
For example, $|x\rangle \rightarrow |3x \bmod 4\rangle$ is the action of $H^2$.

%P2.1.2 - Representation in terms of qubit operations
We use $F$ and $F^\dag$ to rewrite Eq. (\ref{basis_action}) as operations on the underlying qubits.
The qudit Pauli group generators $X$ and $Z$,
\begin{subequations}
\label{qubit_action}
\begin{align}
\vcenter{\vbox{\Qcircuit @C=0.5em @R=0.4em {
  & \qslash & \gate{X}  & \qw \\
}}} &=
\vcenter{\vbox{\Qcircuit @C=0.5em @R=0.4em {
 & \qslash & \multigate{1}{F^\dag}  & \ctrl{1} & \gate{X} & \multigate{1}{F}  & \qslash & \qw \\
 & & \pureghost{F^\dag}            & \targ & \push{\rule{0em}{1.2em}}  \qw & \ghost{F} &
}}} \\
\vcenter{\vbox{\Qcircuit @C=0.5em @R=0.4em {
 & \qslash & \gate{Z}  & \qw \\
}}}  &= 
\vcenter{\vbox{\Qcircuit @C=0.5em @R=0.4em {
 & \qslash & \multigate{1}{F^\dag}  & \gate{S} & \multigate{1}{F}  & \qslash & \qw \\
 & & \pureghost{F^\dag}            & \gate{Z} & \ghost{F} &
}}} \, ,
\end{align}
contain only qubit Clifford operations.
By contrast, the qudit Clifford group generators \textsc{cnot}, $H$, and $S$,
\begin{align}
\vcenter{\vbox{\Qcircuit @C=0.65em @R=1.0em {
 & \qslash & \ctrl{1}  & \qw \\
 & \qslash & \targ  & \qw \\
}}} &= 
\vcenter{\vbox{\Qcircuit @C=0.5em @R=0.4em {
 & \qslash & \multigate{1}{F^\dag}  & \ctrl{2} & \ctrl{2} & \qw    & \multigate{1}{F}  & \qslash & \qw \\
 & & \pureghost{F^\dag}            & \qw      & \qw    & \ctrl{2} & \ghost{F} & \\
 & \qslash & \multigate{1}{F^\dag}  & \ctrl{1} & \targ & \qw       & \multigate{1}{F}  & \qslash & \qw \\
 & & \pureghost{F^\dag}            & \targ    & \qw   & \targ     & \ghost{F} &
}}} \label{qudit_CNOT} \\
\vcenter{\vbox{\Qcircuit @C=0.5em @R=0.4em {
  & \qslash & \gate{H}  & \qw \\
}}}  & = 
\vcenter{\vbox{\Qcircuit @C=0.5em @R=0.4em {
 & \qslash & \multigate{1}{F^\dag}  & \qw & \ctrl{1} & \gate{H} & \qswap \qwx[1] & \multigate{1}{F}  & \qslash & \qw \\
 & & \pureghost{F^\dag}            & \gate{H} & \gate{S} & \qw & \qswap & \ghost{F} &
}}} \label{qudit_H} \\
\vcenter{\vbox{\Qcircuit @C=0.5em @R=0.4em {
 & \qslash & \gate{S}  & \qw \\
}}} & = 
\vcenter{\vbox{\Qcircuit @C=0.5em @R=0.4em {
 & \qslash & \multigate{1}{F^\dag}  & \gate{T} & \ctrl{1} & \multigate{1}{F}  & \qslash & \qw \\
 & & \pureghost{F^\dag}            & \gate{Z} & \gate{Z} & \ghost{F} &
}}} \, ,
\end{align}
\end{subequations}
 contain $T,$ controlled-$S,$ and Toffoli gates, which are standard qubit non-Clifford operations.
If we label the $n$th level of the $m$-dimensional qudit Clifford hierarchy \cite{gate_teleportation} as $C_n^m$,
 then it is clear from Eq. (\ref{qubit_action}) that $C_2^4 \subset C_3^2$.
$C_2^4$ must be a strict subset of $C_3^2$ because qudit Clifford operations are not universal.

%P2.1.3 - Qubit Pauli operators as qudit Clifford operators
We can also use $F$ and $F^\dag$ to rewrite the qubit Pauli group generators as operations on qudits.
Two in particular,
\begin{subequations}
\label{Pauli_in_Pauli}
\begin{align}
\vcenter{\vbox{\Qcircuit @C=0.5em @R=0.4em {
 & \push{\rule{0em}{1.2em}} \qw  & \qw \\
 & \gate{X}  & \qw \\
}}}  &= 
\vcenter{\vbox{\Qcircuit @C=0.5em @R=0.4em {
 & \multigate{1}{F}  & \qslash & \gate{X} & \gate{X} & \multigate{1}{F^\dag}  & \qw \\
 & \ghost{F}            & &    \push{\rule{0em}{1.2em}}            &  & \pureghost{F^\dag} & \qw
}}} \\
\vcenter{\vbox{\Qcircuit @C=0.5em @R=0.4em {
 & \gate{Z}  & \qw \\
 & \push{\rule{0em}{1.2em}} \qw  & \qw \\
}}}  &= 
\vcenter{\vbox{\Qcircuit @C=0.5em @R=0.4em {
 & \multigate{1}{F}  & \qslash &  \gate{Z} & \gate{Z} & \multigate{1}{F^\dag}  & \qw \\
 & \ghost{F}            &  &  \push{\rule{0em}{1.2em}}            & & \pureghost{F^\dag} & \qw
}}} \, ,
\end{align}
\end{subequations}
 are also in the qudit Pauli group.
The other two generators,
\begin{subequations}
\label{Pauli_in_Clifford}
\begin{align}
\vcenter{\vbox{\Qcircuit @C=0.5em @R=0.4em {
 & \gate{X}  & \qw \\
 & \push{\rule{0em}{1.2em}} \qw  & \qw \\
}}}  = 
\vcenter{\vbox{\Qcircuit @C=0.5em @R=0.4em {
 & \multigate{1}{F}  & \qslash &  \gate{H}  & \gate{H} & \gate{X} & \multigate{1}{F^\dag}  & \qw \\
 & \ghost{F}            & &   \push{\rule{0em}{1.2em}}            &    &                & \pureghost{F^\dag} & \qw
}}} \\
\vcenter{\vbox{\Qcircuit @C=0.5em @R=0.4em {
 & \push{\rule{0em}{1.2em}} \qw  & \qw \\
 & \gate{Z}  & \qw \\
}}}  = 
\vcenter{\vbox{\Qcircuit @C=0.5em @R=0.4em {
 & \multigate{1}{F}  & \qslash & \gate{S} & \gate{S} & \gate{Z^\dag} & \multigate{1}{F^\dag}  & \qw \\
 & \ghost{F}            & &  \push{\rule{0em}{1.2em}}               & &               & \pureghost{F^\dag} & \qw
}}} \, ,
\end{align}
\end{subequations}
 are in the qudit Clifford group.
We observe that $C_1^2 \subset C_2^4$.

%P2.1.4 - Qubit Pauli operators in a qudit Pauli basis
We make a final observation about fusion by decomposing the qudit Clifford operators in Eq. (\ref{Pauli_in_Clifford}) in a qudit Pauli basis,
\begin{subequations}
\label{Pauli_algebraic}
\begin{align}
 XH^2 &= \tfrac{1}{2} \left( X + X Z^2 + X^\dag + Z^2 X^\dag \right)   \\
 Z^\dag S^2 &= \tfrac{1}{\sqrt{2}} \left( \omega^* Z + \omega Z^\dag \right) .
\end{align}
\end{subequations}
In the conversion between qubits and qudits, the new Pauli $Z$ operators are functions of the old Pauli $Z$ operators.
This is an asymmetry in the qudit representation since the new Pauli $X$ operators are functions of both old Pauli $X$ and $Z$ operators.
A complementary representation is defined in the appendix.

%S2.2 - Entangle all the representations
\subsection{Hybrid Clifford operations\label{hybrid_Clifford}}

%P2.2.1 - Review previous research on hybrid Clifford operations between qudits of different dimension
A provable construction of all Clifford operations between qudits of different dimensions
 is an open problem \cite{hybrid_Clifford} that we do not attempt to solve here.
We simply introduce additional generators of the Clifford group to entangle qubits and qudits
 that are sufficient for the purpose of this paper.
We conjecture that they are sufficient to generate the entire Clifford group.

%P2.2.2 - Hybrid CNOT circuit elements
The entangling gates that we consider are generalizations of \textsc{cnot} between a qubit and qudit.
They have two orientations,
\begin{subequations}
\label{hybrid_Clifford_ops}
\begin{align}
\vcenter{\vbox{\Qcircuit @C=0.5em @R=0.4em {
 & \qslash & \ctrl{1}  & \push{\rule{0em}{1.2em}} \qw \\
 & \qw & \targ  & \push{\rule{0em}{1.2em}} \qw
}}} & = 
\vcenter{\vbox{\Qcircuit @C=0.5em @R=0.3em {
 & \qslash & \multigate{1}{F^\dag}  & \ctrl{2} & \multigate{1}{F}  & \qslash & \push{\rule{0em}{1.2em}} \qw \\
 &             & \pureghost{F^\dag}     & \qw      & \ghost{F}            & \push{\rule{0em}{1.2em}} \\
 & \qw      & \qw                              & \targ    & \qw                     & \qw & \push{\rule{0em}{1.2em}} \qw
}}} \\
\vcenter{\vbox{\Qcircuit @C=0.5em @R=0.4em {
 & \qw & \ctrl{1} & \push{\rule{0em}{1.2em}} \qw \\
 & \qslash & \targ  & \push{\rule{0em}{1.2em}} \qw \\
}}} & = 
\vcenter{\vbox{\Qcircuit @C=0.5em @R=0.3em  {
 & \qw & \qw                              & \ctrl{2}     & \qw    & \qw & \push{\rule{0em}{1.2em}} \qw  \\
 & \qslash & \multigate{1}{F^\dag}  & \qw & \multigate{1}{F}  & \qslash & \push{\rule{0em}{1.2em}} \qw \\
 & & \pureghost{F^\dag}            & \targ & \ghost{F} & \push{\rule{0em}{1.2em}}
}}} \, ,
\end{align}
\end{subequations}
 which are related by $H$ conjugation on both qubit and qudit.

%P2.2.3 - CNOT operator transformation rules (show for both)
For convenience, we extend the Pauli transformation rules in Eq. (\ref{Clifford_action})
 to include the hybrid \textsc{cnot} operations,
\begin{subequations}
\label{Clifford_action2}
\begin{align}
\vcenter{\vbox{\Qcircuit @C=0.5em @R=0.4em {
 & \qslash & \gate{X} & \ctrl{1} & \qw \\
 & \qw & \push{\rule{0em}{1.2em}} \qw         & \targ    & \qw
}}} &=
\vcenter{\vbox{\Qcircuit @C=0.5em @R=0.4em {
 & \qslash  & \ctrl{1} & \gate{X} & \qw \\
 & \qw & \targ &  \gate{X} & \qw
}}} \\
\vcenter{\vbox{\Qcircuit @C=0.5em @R=0.4em {
 & \qslash & \push{\rule{0em}{1.2em}} \qw & \ctrl{1} & \qw \\
 & \qw & \gate{X} & \targ    & \qw
}}} &=
\vcenter{\vbox{\Qcircuit @C=0.5em @R=0.4em {
 & \qslash  & \ctrl{1} & \push{\rule{0em}{1.2em}} \qw & \qw \\
 & \qw & \targ &  \gate{X} & \qw
}}} \\
\vcenter{\vbox{\Qcircuit @C=0.5em @R=0.4em {
 & \qslash & \gate{Z} & \ctrl{1} & \qw \\
 & \qw & \push{\rule{0em}{1.2em}} \qw         & \targ    & \qw
}}} &=
\vcenter{\vbox{\Qcircuit @C=0.5em @R=0.4em {
 & \qslash  & \ctrl{1} & \gate{Z} & \qw \\
 & \qw & \targ & \push{\rule{0em}{1.2em}} \qw & \qw
}}} \\
\vcenter{\vbox{\Qcircuit @C=0.5em @R=0.4em {
 & \qslash & \push{\rule{0em}{1.4em}} \qw & \ctrl{1} & \qw \\
 & \qw & \gate{Z} & \targ    & \qw
}}} &=
\vcenter{\vbox{\Qcircuit @C=0.5em @R=0.4em {
 & \qslash  & \ctrl{1} & \gate{Z^2} & \qw \\
 & \qw & \targ &  \gate{Z} & \qw
}}} \\
\vcenter{\vbox{\Qcircuit @C=0.5em @R=0.5em {
 & \qw & \gate{X} & \ctrl{1} & \qw \\
 & \qslash & \push{\rule{0em}{1.2em}} \qw         & \targ    & \qw
}}} &=
\vcenter{\vbox{\Qcircuit @C=0.5em @R=0.4em {
 & \qw  & \ctrl{1} & \gate{X} & \qw \\
 & \qslash & \targ &  \gate{X^2} & \qw
}}} \\
\vcenter{\vbox{\Qcircuit @C=0.5em @R=0.4em {
 & \qw & \push{\rule{0em}{1.2em}} \qw & \ctrl{1} & \qw \\
 & \qslash & \gate{X} & \targ    & \qw
}}} &=
\vcenter{\vbox{\Qcircuit @C=0.5em @R=0.4em {
 & \qw  & \ctrl{1} & \push{\rule{0em}{1.2em}} \qw & \qw \\
 & \qslash & \targ &  \gate{X} & \qw
}}} \\
\vcenter{\vbox{\Qcircuit @C=0.5em @R=0.4em {
 & \qw & \gate{Z} & \ctrl{1} & \qw \\
 & \qslash & \push{\rule{0em}{1.2em}} \qw         & \targ    & \qw
}}} &=
\vcenter{\vbox{\Qcircuit @C=0.5em @R=0.4em {
 & \qw  & \ctrl{1} & \gate{Z} & \qw \\
 & \qslash & \targ & \push{\rule{0em}{1.2em}} \qw & \qw
}}} \\
\vcenter{\vbox{\Qcircuit @C=0.5em @R=0.4em {
 & \qw & \push{\rule{0em}{1.3em}} \qw & \ctrl{1} & \qw \\
 & \qslash & \gate{Z} & \targ    & \qw
}}} &=
\vcenter{\vbox{\Qcircuit @C=0.5em @R=0.4em {
 & \qw  & \ctrl{1} & \gate{Z} & \qw \\
 & \qslash & \targ &  \gate{Z} & \qw
}}} \, .
\end{align}
\end{subequations}
By combining Eqs. (\ref{Clifford_action}) and (\ref{Clifford_action2}) with standard transformation rules for qubits,
 we can propagate Pauli operators through any stabilizer circuit containing both qubits and qudits.

%S3
\section{Conversion circuits}

%P3.1 - Resource state
The qudit resource state for both fusion and fission is
\begin{equation}
 | F \rangle = \tfrac{1}{\sqrt{2}} \left( |0\rangle + | 1 \rangle \right) ,
\end{equation}
 which is the fusion of a simple qubit stabilizer state
\begin{equation} \label{define_F}
\vcenter{\vbox{\Qcircuit @C=0.5em @R=0.8em {
 \lstick{\ket{0}} & \gate{H} & \multigate{1}{F} & \qslash &  \rstick{\ket{F}} \qw \\
 \lstick{\ket{0}} & \qw         & \ghost{F}           &
 }}} \ \ \ \ \ \ \ \ .
\end{equation}
Because $| F \rangle$ is a resource state for non-Clifford operations,
 it should be expected that a stabilizer circuit implementation of Eq. (\ref{define_F})
 merely teleports an ancilla qudit prepared as $| F \rangle$. 

%P3.2 - Partial fusion in relation to complete fusion
It is convenient to define partial operations for fusion and fission
that either prepare an input qubit in a predetermined state
 or measure an output qubit in a predetermined basis,
\begin{subequations}
\label{partial_ops}
\begin{align}
\vcenter{\vbox{\Qcircuit @C=0.5em @R=0.8em {
  & \multigate{1}{F}  & \qslash & \qw \\
  & \pureghost{F} \cw &
}}} &=
\vcenter{\vbox{\Qcircuit @C=0.5em @R=0.4em {
  &  \push{\rule{0em}{1.2em}} \qw & \qw &  \qw                      & \qw                  & \multigate{1}{F}  & \qslash & \qw \\
  &  \push{\rule{0.8em}{0em}\rule{0em}{1.2em}} & \lstick{|0\rangle} & \gate{H} & \gate{Z} \cwx[1] & \ghost{F}  & \\
 &  \push{\rule{0em}{1.2em}} \cw & \cw & \cw & \control \cw  
}}} \\
\vcenter{\vbox{\Qcircuit @C=0.5em @R=0.8em {
  & \pureghost{F}  \cw & \qslash & \qw \\
  & \multigate{-1}{F}  &
}}} &=
\vcenter{\vbox{\Qcircuit @C=0.5em @R=0.4em {
 &  \push{\rule{0em}{1.2em}} \cw &  \cw                     & \control \cw   \\
  &  \push{\rule{0.8em}{0em}\rule{0em}{1.2em}} & \lstick{|0\rangle} & \gate{X} \cwx[-1] & \ghost{F}  & \qslash & \qw \\
  &  \push{\rule{0em}{1.2em}} \qw & \qw &  \qw                      & \multigate{-1}{F}  & 
}}} \\
\vcenter{\vbox{\Qcircuit @C=0.5em @R=0.8em {
  & \qslash & \multigate{1}{F^\dag}  & \qw \\
   &  & \pureghost{F^\dag} & \cw    
}}} &=
\vcenter{\vbox{\Qcircuit @C=0.5em @R=0.4em {
 &  \qslash & \multigate{1}{F^\dag}  & \qw &  \qw & \push{\rule{0em}{1.2em}} \qw \\
  &  & \pureghost{F^\dag} & \gate{H} &  \meter & \push{\rule{0em}{1.2em}} \cw    
}}}  \\
\vcenter{\vbox{\Qcircuit @C=0.5em @R=0.8em {
 &  \qslash & \multigate{1}{F^\dag}  & \cw \\
  &  & \pureghost{F^\dag} & \qw    
}}} &=
\vcenter{\vbox{\Qcircuit @C=0.5em @R=0.4em {
 &  \qslash & \multigate{1}{F^\dag}  & \meter{Z} & \push{\rule{0em}{1.2em}} \cw \\
  &  & \pureghost{F^\dag} & \qw &  \push{\rule{0em}{1.2em}} \qw    
}}} \, .
\end{align}
\end{subequations}
A complete set of quantum states can be fused or split by these operations, but full quantum coherence is not preserved.

%P3.3 - Stabilizer circuit implementation of partial fusion and fission
An advantage of the partial fusion and fission operations in Eq. (\ref{partial_ops}) is that
 they have stabilizer circuit implementations,
\begin{subequations}
\label{F_partial}
\begin{align}
\vcenter{\vbox{\Qcircuit @C=0.5em @R=0.8em {
  & \multigate{1}{F}  & \qslash & \qw \\
  & \pureghost{F}    \cw       & &
}}} &=
\vcenter{\vbox{\Qcircuit @C=0.5em @R=0.4em {
  & \push{\rule{0em}{1.2em}} \qw                      & \qw                 & \qw              & \qw        & \targ             & \meter \cwx[1] \\
  & \push{\rule{0.8em}{0em}\rule{0em}{1.2em}} & \lstick{\ket{0}} & \qslash \qw & \gate{H} & \ctrl{-1}         & \gate{XH^2}  & \gate{Z^\dag S^2} & \qw \\
  & \push{\rule{0em}{1.2em}} \cw                      & \cw                  & \cw             & \cw         & \cw               & \cw                & \control  \cw  \cwx[-1] 
}}} \\
\vcenter{\vbox{\Qcircuit @C=0.5em @R=0.8em {
  & \pureghost{F}  \cw & \qslash & \qw \\
  & \multigate{-1}{F}  &
}}} &=
\vcenter{\vbox{\Qcircuit @C=0.5em @R=0.4em {
  & \push{\rule{0em}{1.2em}} \cw                        &  \cw                 & \cw             &\cw       &  \control \cw \cwx[1] \\
  &  \push{\rule{0.8em}{0em}\rule{0em}{1.2em}} & \lstick{\ket{0}} & \qslash \qw & \targ    &  \gate{XH^2}            & \gate{Z^\dag S^2}     & \qw \\
  & \push{\rule{0em}{1.2em}} \qw                       & \qw                  & \qw             & \ctrl{-1} & \gate{H}                  & \meter \cwx[-1]
}}} \\
\vcenter{\vbox{\Qcircuit @C=0.5em @R=0.8em {
 & \qslash & \multigate{1}{F^\dag}  &  \qw \\
 &   & \pureghost{F^\dag} &   \cw    
}}} &= 
\vcenter{\vbox{\Qcircuit @C=0.5em @R=0.4em {
  &\push{\rule{0.8em}{0em}\rule{0em}{1.2em}}  & \lstick{\ket{0}} & \qw         & \qw        & \targ       & \qw                        & \qw         & \gate{S} & \qw \\
 & \qslash  \push{\rule{0em}{1.2em}}                 & \qw                  & \qw         & \targ      & \ctrl{-1}   & \gate{Z^\dag S^2} & \gate{H} & \meter \cwx[-1] & \\
 & \push{\rule{0.8em}{0em}\rule{0em}{1.2em}}  &  \lstick{\ket{0}} & \gate{H} & \ctrl{-1}  & \gate{H} &  \meter \cwx[-1]     & \cw         &\cw          & \cw
}}} \\
\vcenter{\vbox{\Qcircuit @C=0.5em @R=0.8em {
  & \qslash & \multigate{1}{F^\dag}  &  \cw \\
  &  & \pureghost{F^\dag} &   \qw    
}}} &=
\vcenter{\vbox{\Qcircuit @C=0.5em @R=0.4em {
 &  \push{\rule{0.8em}{0em}\rule{0em}{1.2em}}   & \lstick{\ket{0}} & \qw        & \qw       & \targ     & \meter \cwx[1] &  \cw                 & \cw \\
 &  \qslash \push{\rule{0em}{1.2em}}                   & \qw                  & \qw        & \targ     & \ctrl{-1} & \gate{XH^2}    & \meter \cwx[1] \\
 &   \push{\rule{0.8em}{0em}\rule{0em}{1.2em}} &  \lstick{\ket{0}} & \gate{H} & \ctrl{-1} & \qw      &  \qw                 & \gate{HSH}      & \qw
}}} \, .
\end{align}
\end{subequations}
These are all standard quantum circuits for state teleportation partially rewritten using qudit Clifford operations.

%P3.4 - Complete fusion and fission circuits
With access to an ancilla qudit initialized to $|F\rangle$,
 stabilizer circuits for complete fusion and fission are
\begin{subequations}
\begin{align}
\vcenter{\vbox{\Qcircuit @C=0.5em @R=0.8em {
  & \multigate{1}{F}  & \qslash & \qw \\
  & \ghost{F} &
}}}  & =
\vcenter{\vbox{\Qcircuit @C=0.5em @R=0.4em {
  & \qw & \qw                & \qw & \qw      & \targ     & \meter \cwx[1] \\
  & \push{\rule{0.8em}{0em}\rule{0em}{1.2em}} & \lstick{\ket{F}} & \qslash & \targ     & \ctrl{-1} & \gate{XH^2 } &  \gate{Z^\dag S^2 } & \qw \\
  & \qw & \qw                & \qw & \ctrl{-1} & \qw      & \gate{H}                &  \meter \cwx[-1]
}}} \\
\vcenter{\vbox{\Qcircuit @C=0.5em @R=0.8em {
  & \qslash & \multigate{1}{F^\dag}  &  \qw \\
   &                           & \pureghost{F^\dag}             &   \qw    
}}}   & =
\vcenter{\vbox{\Qcircuit @C=0.5em @R=0.4em {
 & \qslash  \push{\rule{0em}{1.2em}} & \qw               & \qw  & \ctrl{2}              & \multigate{1}{F^\dag}              & \ctrl{3} & \qw & \qw \\
  &   \push{\rule{0.8em}{0em}\rule{0em}{1.6em}}  &             &                 &                         & \pureghost{F^\dag}  & \cw & \control \cw \cwx[2] \\
  & \push{\rule{0.8em}{0em}\rule{0em}{1.2em}} &  \lstick{\ket{F}}  & \qslash & \gate{X^\dag}   &  \multigate{1}{F^\dag} & \control \cw \\
  &                         &                           &                           &  &  \pureghost{F^\dag}             & \targ & \gate{Z} & \qw
}}} \, . \label{complete_fission}
\end{align}
\end{subequations}
Again, these are standard circuits in nonstandard notation.

%F1 - Error detection circuit
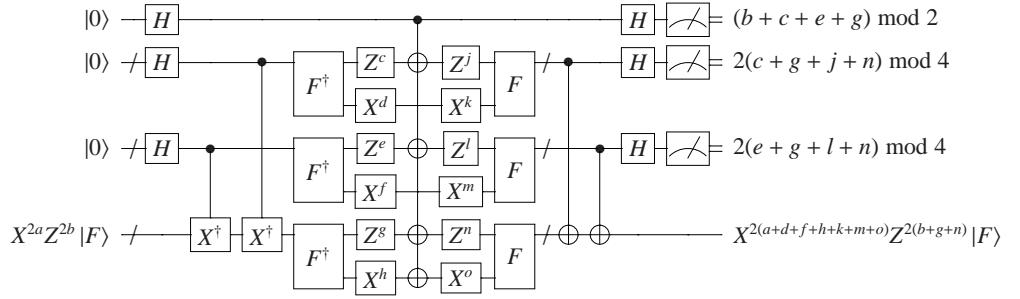
\begin{figure*}[t]
$\vcenter{\vbox{\Qcircuit @C=0.5em @R=0.4em {
  \lstick{\ket{0}}                            & \qw        & \gate{H} & \qw                    & \qw                   & \qw                                & \qw            & \ctrl{6} & \qw            & \qw                      & \qw        & \qw      & \qw      & \gate{H} & \meter & \rstick{(b+c+e+g) \bmod 2} \cw \\
  \lstick{\ket{0}}                            & \qslash & \gate{H} & \qw                    & \ctrl{4}              & \multigate{1}{F^\dag} & \gate{Z^c} & \targ    & \gate{Z^j} & \multigate{1}{F} & \qslash & \ctrl{4} & \qw      & \gate{H} & \meter & \rstick{2(c+g+j+n) \bmod 4} \cw \\
                                                      &               &                 &                           &                           & \pureghost{F^\dag}    & \gate{X^d} & \qw     & \gate{X^k} & \ghost{F}    \\
  \lstick{\ket{0}}                            & \qslash & \gate{H}  & \ctrl{2}              & \qw                   & \multigate{1}{F^\dag}  & \gate{Z^e} & \targ   & \gate{Z^l} & \multigate{1}{F} & \qslash & \qw     & \ctrl{2} & \gate{H} & \meter & \rstick{2(e+g+l+n) \bmod 4} \cw \\
                                                      &               &                  &                          &                           & \pureghost{F^\dag}     & \gate{X^f}  & \qw    & \gate{X^m}  & \ghost{F}    \\
  \lstick{X^{2a} Z^{2b} \ket{F}}  & \qslash & \qw           & \gate{X^\dag} & \gate{X^\dag} & \multigate{1}{F^\dag}  & \gate{Z^g} & \targ  & \gate{Z^n} & \multigate{1}{F} & \qslash & \targ   & \targ    & \qw          & \qw      & \rstick{X^{2(a+d+f+h+k+m+o)} Z^{2(b+g+n)} \ket{F}} \qw \\
                                                      &               &                  &                           &                          & \pureghost{F^\dag}     & \gate{X^h} & \targ  & \gate{X^o} & \ghost{F} 
}}}$
\caption{\label{Z_detect} $Z^2$ error detection circuit for twirled noisy $|F\rangle$ states with all possible error locations included.
}
\end{figure*}

%S4
\section{Resource states\label{resource}}

%P4.1 - Resource state conversion
Although $|F\rangle$ is an unconventional resource state, it can be used to extract the non-Clifford operations in Eq. (\ref{qubit_action})
 into the conventional resource states of $C_3^2$ gate teleportation \cite{gate_teleportation},
\begin{subequations} \label{compile_all}
\begin{align}
\vcenter{\vbox{\Qcircuit @C=0.5em @R=0.6em {
  \lstick{\ket{0}} & \gate{H} & \gate{T} & \qw
}}} &= \ \ \ \ \ \ \ \
\vcenter{\vbox{\Qcircuit @C=0.5em @R=0.6em {
  \lstick{\ket{F}} & \qslash & \gate{S} & \multigate{1}{F^\dag} & \qw \\
                         &             &               & \pureghost{F^\dag}    &
}}} \label{compile_T} \\
\vcenter{\vbox{\Qcircuit @C=0.5em @R=0.6em {
  \lstick{\ket{0}} & \gate{H} & \ctrl{1} & \push{\rule{0em}{1.2em}} \qw \\
  \lstick{\ket{0}} & \gate{H} & \gate{S} & \qw
}}} &= \ \ \ \ \ \ \ \
\vcenter{\vbox{\Qcircuit @C=0.5em @R=0.6em {
  \lstick{\ket{F}} & \qslash & \ctrl{2}   & \multigate{1}{F^\dag} & \qw \\
                         &             &               & \pureghost{F^\dag}    & \\
  \lstick{\ket{F}} & \qslash & \gate{Z} & \multigate{1}{F^\dag} & \qw \\
                         &             &               & \pureghost{F^\dag}    &
}}} \label{compile_CS} \\
\vcenter{\vbox{\Qcircuit @C=0.5em @R=0.6em {
  \lstick{\ket{0}} & \gate{H} & \ctrl{1} & \push{\rule{0em}{1.2em}} \qw \\
  \lstick{\ket{0}} & \gate{H} & \ctrl{1} & \push{\rule{0em}{1.2em}} \qw \\
  \lstick{\ket{0}} & \qw        & \targ & \qw
}}} &= \ \ \ \ \ \ \ \
\vcenter{\vbox{\Qcircuit @C=0.5em @R=0.6em {
  \lstick{\ket{F}} & \qslash & \ctrl{2}   & \multigate{1}{F^\dag} & \qw \\
                         &             &               & \pureghost{F^\dag}    & \\
  \lstick{\ket{F}} & \qslash & \gate{X^\dag} & \multigate{1}{F^\dag} & \qw \\
                         &             &               & \pureghost{F^\dag}    & \qw
}}} \, . \label{compile_CCX}
\end{align}
\end{subequations}
The $T$, controlled-$S$, and Toffoli gates require one, two, and three copies of $|F\rangle$ to implement respectively.
Whether or not $|F\rangle$ can be prepared from finite numbers of these conventional resource states is unclear and left to future work.

%P4.2 - Summary of state preparation
Fault-tolerant state preparation of $|F\rangle$ is similar to standard magic state distillation \cite{distillation}.
Faulty $|F\rangle$ states are stochastically twirled using the stabilizers of $|F\rangle$ in Eq. (\ref{Pauli_in_Clifford}),
\begin{subequations}
\begin{align}
 W_X(\rho) &= \tfrac{1}{2} \left( \rho + X H^2 \rho H^2 X^\dag  \right) \\
 W_Z(\rho) &= \tfrac{1}{2} \left( \rho + Z^\dag S^2 \rho S^2 Z  \right),
\end{align}
\end{subequations}
 which reduces errors to a statistical mixture of $X^2$ and $Z^2$,
\begin{align} \label{rhoF}
 \rho_F(p_X,p_Z,p_{XZ}) &= (1 - p_X - p_Z - p_{XZ}) | F \rangle \langle F | \notag \\
  & \ \ \ + p_X X^2 | F \rangle \langle F | X^2 + p_Z Z^2 | F \rangle \langle F |Z^2 \notag \\
  & \ \ \ + p_{XZ} X^2 Z^2 | F \rangle \langle F | Z^2 X^2 .
\end{align}
These error probabilities are then reduced by applying error detection stabilizer circuits to multiple copies of $\rho_F$.

%P4.3 - Error detection circuits
The detection circuit for $X^2$ errors is similar to Eq. (\ref{complete_fission}),
\begin{equation} \label{X_detect}
\vcenter{\vbox{\Qcircuit @C=0.5em @R=0.4em {
  \lstick{X^{2a} Z^{2b} \ket{F}} & \qslash & \ctrl{1} &  \qw & \ctrl{2} & \push{\rule{0em}{1.2em}} \qw & \rstick{X^{2a} Z^{2(b+d)} \ket{F}} \qw \\
  \lstick{X^{2c} Z^{2d} \ket{F}}   & \qslash & \gate{X^\dag}   &  \multigate{1}{F^\dag} & \control \cw \\
  &                        &           &            \pureghost{F^\dag} & \targ & \meter & \rstick{(a + c) \bmod 2} \cw
}}} \ \ \ \ \ \ \ \ \ \ \
\end{equation}
 and consumes one $|F\rangle$ to detect an error in the output $|F\rangle$.
On the underlying qubits, this circuit projectively measures $Z$ on the second qubit
 with a \textsc{cnot} and measurement on the fourth qubit.
It is more complicated as a qudit stabilizer circuit, with
 a partial fission and classically-controlled \textsc{cnot} that cancel a Toffoli gate.
We are unable to find a complementary circuit to detect $Z^2$ errors.
Instead, we use a repetition code to encode a qubit $X$ measurement in Fig. \ref{Z_detect}, which consumes six $|F\rangle$.

%P4.4 - Leading-order error analysis
With statistically equivalent inputs of the form in Eq. (\ref{rhoF})
 and to leading order in $p_X$, $p_Z$, and $p_{XZ}$,
 Eq. (\ref{X_detect}) and Fig. \ref{Z_detect} detect an error in $\rho_F$ as a nonzero measurement outcome
 with probability $2 p_X + 2 p_{XZ}$ and $7 p_Z + 7 p_{XZ}$ or otherwise output
\begin{subequations}
\begin{align}
 2 \times \rho_F &\rightarrow \rho_F(p_X^2 + p_{XZ}^2, 2 p_Z, 2 p_X p_{XZ}) \\
 7 \times \rho_F &\rightarrow \rho_F( 7 p_X, 6 p_Z^3 + 18 p_Z p_{XZ}^2, 6 p_{XZ}^3 + 18 p_Z^2 p_{XZ})
\end{align}
\end{subequations}
 respectively.
For uniform error reduction, an incommensurate nesting of $X^2$ and $Z^2$ detection circuits is required.
A greedy nesting that always suppresses the most probable error has a threshold of $p \approx 0.17$ for $p_X = p_Z = (1-p)p$ and $p_{XZ} = p^2$.

%P4.5 - Comparison to other methods for resource state preparation
We have established that fault-tolerant $|F\rangle$ state preparation is possible with stabilizer circuits,
 but our construction is not efficient compared to state-of-the-art magic state distillation.
Our input-output ratio for quadratic error reduction is $\approx 6.8$, 
 which is inferior to 4 for five-dimensional qudit magic states \cite{qudit_magic_distillation}
 and 2 for large numbers of qubit magic states \cite{qubit_asymptotic_distillation}.
Future research should search for new $|F\rangle$ distillation protocols and a
 direct implementation of logical $F$ and $F^\dag$ gates by converting
 between two qubit surface codes and a qudit surface code.

%S5
\section{Conclusions}

%P5.1 - Compilation advantages
While qubit fusion is not an efficient method for universal quantum computation in its present form,
 Clifford+$F$ circuits have an advantage relative to Clifford+$T$ circuits.
The cost of a quantum computation compiled into a Clifford+$T$ circuit
 is often measured in the number and depth of $T$ gates \cite{count_and_depth}.
We can recompile Clifford+$T$ circuits into Clifford+$F$ circuits by replacing each $T$ gate with an $F$ gate using Eq. (\ref{compile_all}).
We can reduce the number of $F$ gates by optimizing circuits to
 use two $F$ gates instead of three $T$ gates \cite{quantum_circuit} for each controlled-$S$ gate
 and three $F$ gates instead of four $T$ gates for each Toffoli gate \cite{Toffoli_compile}.
Thus Clifford+$F$ circuits can be more efficient than Clifford+$T$ circuits
 in their usage of basic resource states.

%P5.2 - Future generalizations
There are two natural generalizations of qubit fusion.
The first generalization is to \textit{qudit} fusion,
 where a $p$-dimensional qudit and a $q$-dimensional qudit merge into a $pq$-dimensional qudit.
Using the standard qudit Pauli and Clifford groups \cite{qudit_stabilizer},
 we can attempt to generalize all the quantum circuit identities in this paper.
Of particular note is the recursive construction of $H$ in Eq. (\ref{qudit_H}).
$H$ is a quantum Fourier transform (QFT), and its decomposition into lower-dimensional QFTs, phase gates, and data permutations
 is very similar to the Cooley-Tukey fast Fourier transform algorithm \cite{FFT}.
The second generalization is to other pairs of Pauli and Clifford groups that operate on the same Hilbert space and have similar nesting structure.
We require each Clifford group to contain both Pauli groups and have elements that are not contained within the other Clifford group.
Qubit fusion results from the use of qubit and qudit Pauli groups,
 and magic state distillation results from the use of two qubit Pauli groups generated by $\{X,Z\}$ and $\{XS,Z\}$ \cite{distillation}.
Other sets of quantum operations with similar group structure
 might also be relevant for universal quantum computation.

%P5.3 - Putting it all together
Ultimately, we must judge schemes for universal quantum computation holistically.
Their relative value will depend on ease of implementation on physical qubits,
 compatibility with quantum error correction,
 the threshold and code efficiency of compatible codes,
 an efficient method for circuit compilation,
 and efficient resource distillation or other implementation of a logical non-Clifford gate.
These issues have been considered extensively for Clifford+$T$ circuits on qubits \cite{UQC_plans},
 but there are inefficiencies \cite{UQC_criticism}
 that warrant the consideration of other schemes, such as Clifford+$F$ circuits on qubits and qudits.

\begin{acknowledgments}
This work was supported by the Laboratory Directed Research and Development program at Sandia National Laboratories.
Sandia National Laboratories is a multi-program laboratory managed and 
operated by Sandia Corporation, a wholly owned subsidiary of Lockheed 
Martin Corporation, for the U.S. Department of Energy's National  
Nuclear Security Administration under contract DE-AC04-94AL85000.
\end{acknowledgments}

\appendix*

%SA - Restore the broken symmetry
\section{Alternate qudit representation\label{alternate_qudit}}

%PA.1 - Motivation for an alternate representation
It is interesting to consider other qudit representations that complement Section \ref{standard_qudit}
 with a qubit to qudit conversion that preserves Eq. (\ref{Pauli_in_Pauli})
 and switches the role of $X$ and $Z$ in Eq. (\ref{Pauli_algebraic}).
We satisfy these constraints with an alternate fusion gate $G$,
\begin{equation}
\label{G_define}
\vcenter{\vbox{\Qcircuit @C=0.5em @R=0.6em {
 & \multigate{1}{G} & \qslash & \qw \\
 & \ghost{G}             &              &
}}} =
\vcenter{\vbox{\Qcircuit @C=0.5em @R=0.35em {
 & \qswap \qwx[1] & \gate{H} & \multigate{1}{F} & \qslash & \gate{H^\dag} & \qw \\
 & \qswap & \gate{H} & \ghost{F}             &              &
}}} \, .
\end{equation}
$F$ and $G$ are equivalent up to stabilizer operations.
All results obtained in this paper for $F$ gates and $|F\rangle$ states generalize to this qudit representation with minor modifications of circuits.
The computational basis is preserved by $F$ in Eq. (\ref{fusion_and_fission})
 because it maps between $Z$ eigenstates of qubits and qudits.
$G$ has the complementary effect of mapping between $X$ eigenstates.

%PA.2 - Change to qudit Pauli and Clifford operations
Complementary to Eq. (\ref{qubit_action}), we rewrite Eq. (\ref{basis_action})
 using $G$ and $G^\dag$ as a different set of operations on the underlying qubits,
\begin{subequations}
\label{qubit_action2}
\begin{align}
\vcenter{\vbox{\Qcircuit @C=0.5em @R=0.4em {
  & \qslash & \gate{X}  & \qw \\
}}} &=
\vcenter{\vbox{\Qcircuit @C=0.5em @R=0.4em {
 & \qslash & \multigate{1}{G^\dag}  & \qw & \gate{X} & \qw & \multigate{1}{G}  & \qslash & \qw \\
 & & \pureghost{G^\dag}                 & \gate{H} & \gate{S} & \gate{H} & \ghost{G} &
}}} \\
\vcenter{\vbox{\Qcircuit @C=0.5em @R=0.4em {
 & \qslash & \gate{Z}  & \qw \\
}}}  &= 
\vcenter{\vbox{\Qcircuit @C=0.5em @R=0.4em {
 & \qslash & \multigate{1}{G^\dag}  & \push{\rule{0em}{1.2em}} \qw & \ctrl{1} & \multigate{1}{G}  & \qslash & \qw \\
 & & \pureghost{G^\dag}            & \gate{Z} & \targ & \ghost{G} &
}}} \\
\vcenter{\vbox{\Qcircuit @C=0.65em @R=1.0em {
 & \qslash & \ctrl{1}  & \qw \\
 & \qslash & \targ  & \qw \\
}}} &= 
\vcenter{\vbox{\Qcircuit @C=0.5em @R=0.3em {
 & \qslash & \multigate{1}{G^\dag}  & \ctrl{2}     & \qw      & \qw          & \ctrl{1} & \qw         & \multigate{1}{G}  & \qslash & \qw \\
 & & \pureghost{G^\dag}                   & \qw          & \ctrl{2}  & \gate{H} & \ctrl{2} & \gate{H} & \ghost{G} & \\
 & \qslash & \multigate{1}{G^\dag}  & \targ        & \qw       & \qw         & \qw       & \qw         & \multigate{1}{G}  & \qslash & \qw \\
 & & \pureghost{G^\dag}                   & \qw          & \targ     & \qw         & \targ      & \qw         & \ghost{G} &
}}} \\
\vcenter{\vbox{\Qcircuit @C=0.5em @R=0.4em {
  & \qslash & \gate{H}  & \qw \\
}}}  & = 
\vcenter{\vbox{\Qcircuit @C=0.5em @R=0.4em {
 & \qslash & \multigate{1}{G^\dag}  & \qw & \ctrl{1} & \gate{H} & \qswap \qwx[1] & \multigate{1}{G} & \qslash & \qw \\
 & & \pureghost{G^\dag}            & \gate{H} & \gate{S} & \qw & \qswap & \ghost{G} &
}}} \\
\vcenter{\vbox{\Qcircuit @C=0.5em @R=0.4em {
 & \qslash & \gate{S}  & \qw \\
}}} & = 
\vcenter{\vbox{\Qcircuit @C=0.5em @R=0.4em {
 & \qslash & \multigate{1}{G^\dag}  & \gate{T} & \ctrl{1} & \multigate{1}{G} & \qslash & \qw \\
 & & \pureghost{G^\dag}            & \gate{Z} & \gate{Z} & \ghost{G} &
}}} \, .
\end{align}
\end{subequations}
It is now $Z$ rather than $X$ that entangles the underlying qubits.
The qubit operations that implement $H$ and $S$ are unchanged from Eq. (\ref{qubit_action}),
 therefore $H$ and $S$ commute with $F G^\dag$.
 $F G^\dag$ is effectively a qudit non-Clifford operation.
The hybrid Clifford operations of this representation are identical to Eq. (\ref{hybrid_Clifford_ops}).

%PA.3 - Change to qubit Pauli operators
By design, the mapping in Eq. (\ref{Pauli_in_Pauli}) is unchanged for $G$,
\begin{subequations}
\label{Pauli_in_Pauli2}
\begin{align}
\vcenter{\vbox{\Qcircuit @C=0.5em @R=0.4em {
 & \push{\rule{0em}{1.2em}} \qw  & \qw \\
 & \gate{X}  & \qw \\
}}}  &= 
\vcenter{\vbox{\Qcircuit @C=0.5em @R=0.4em {
 & \multigate{1}{G}  & \qslash & \gate{X} & \gate{X} & \multigate{1}{G^\dag}  & \qw \\
 & \ghost{G}            & &    \push{\rule{0em}{1.2em}}            &  & \pureghost{G^\dag} & \qw
}}} \\
\vcenter{\vbox{\Qcircuit @C=0.5em @R=0.4em {
 & \gate{Z}  & \qw \\
 & \push{\rule{0em}{1.2em}} \qw  & \qw \\
}}}  &= 
\vcenter{\vbox{\Qcircuit @C=0.5em @R=0.4em {
 & \multigate{1}{G}  & \qslash &  \gate{Z} & \gate{Z} & \multigate{1}{G^\dag}  & \qw \\
 & \ghost{G}            &  &  \push{\rule{0em}{1.2em}}            & & \pureghost{G^\dag} & \qw
}}} \, .
\end{align}
\end{subequations}
Also by design, the mapping in Eq. (\ref{Pauli_in_Clifford}) is reversed for $G$,
\begin{subequations}
\begin{align}
\vcenter{\vbox{\Qcircuit @C=0.5em @R=0.45em {
 & \gate{X}  & \qw \\
 &    \push{\rule{0em}{1.2em}} \qw  & \qw \\
}}} &=
\vcenter{\vbox{\Qcircuit @C=0.5em @R=0.4em {
 & \multigate{1}{G}  & \qslash & \gate{H} & \gate{S} & \gate{S} & \gate{Z^\dag} & \gate{H^\dag} & \multigate{1}{G^\dag}  & \qw \\
 & \ghost{G}            &    \push{\rule{0em}{1.2em}} &                            &                 &                &                &                 & \pureghost{G^\dag} & \qw
}}} \\
\vcenter{\vbox{\Qcircuit @C=0.5em @R=0.45em {
 &    \push{\rule{0em}{1.2em}} \qw  & \qw \\
 & \gate{Z}  & \qw \\
}}} &=
\vcenter{\vbox{\Qcircuit @C=0.5em @R=0.4em {
 & \multigate{1}{G}  & \qslash &  \gate{H}  & \gate{H} & \gate{Z^\dag} & \multigate{1}{G^\dag}  & \qw \\
 & \ghost{G}            &    \push{\rule{0em}{1.2em}}             &  &    &                & \pureghost{G^\dag} & \qw
}}} \, .
\end{align}
\end{subequations}
Decomposition of these operators in the qudit Pauli basis,
\begin{subequations}
\begin{align}
 H^\dag Z^\dag S^2 H &= \tfrac{1}{\sqrt{2}} \left( \omega^* X + \omega X^\dag \right) \\
 Z^\dag H^2 &= \tfrac{1}{2} \left( Z + X^2 Z + Z^\dag + Z^\dag X^2 \right) ,
\end{align}
\end{subequations}
 clearly demonstrates the complementarity with Eq. (\ref{Pauli_algebraic}).

%PA.4 - Resource state preparation
We can define $|G\rangle$ analogous to $|F\rangle$ in Eq. (\ref{define_F}), but the two states only differ by $H$.
Because of the $H$ and \textsc{swap} on qubits in Eq. (\ref{G_define}), the nature of the errors in a faulty $|G\rangle$ state are opposite that of $|F\rangle$.
$Z^2$ errors become efficient to detect as in Eq. (\ref{X_detect}) and $X^2$ errors become difficult to detect as in Fig. \ref{Z_detect}.
If we could switch representations and preserve the identity of errors, then $|F\rangle$ state preparation would be more efficient.
This might be possible for a logical qudit encoded in a topological code,
 where the local identity of physical errors is decoupled from the global identity of logical errors.
A logical string can be $X$-type in one spatial region and $Z$-type in another.

\end{document}